\RequirePackage{fix-cm}
\documentclass[smallextended]{svjour3}       % onecolumn (second format)
\smartqed  % flush right qed marks, e.g. at end of proof
\usepackage{graphicx}
\usepackage{mathptmx}      % use Times fonts if available on your TeX system
\usepackage{amssymb, amsmath}
\usepackage{tikz}
\usepackage[ruled,vlined,linesnumbered]{algorithm2e}

\newcommand{\mintb}{\textbf{MINTB}}

\newcommand{\fpt}{\textbf{FPT}}
\newcommand{\w}{\textbf{W[2]}}
\newcommand{\np}{\textbf{NP}}
\newcommand{\cnf}{\textbf{CNF}}
\newcommand{\wcnf}{\textbf{weighted CNF SAT}}
\newcommand{\cp}{clause player}
\newcommand{\op}{occurrence player}
\newcommand{\vp}{variable player}

\begin{document}

\title{The Minimum Tollbooth Problem in \\Atomic Network Congestion Games with Unsplittable Flows}
%\subtitle{Do you have a subtitle?\\ If so, write it here}

\titlerunning{MINTB in Atomic Network Congestion Games with Unsplittable Flows}        % if too long for running head

\author{Julian Nickerl
}

%\authorrunning{Short form of author list} % if too long for running head

\institute{Julian Nickerl \at
              University of Ulm \\
              Tel.: +49 (0)731 50 24256\\
              Fax: +49 (0)731 50 1224101\\
              \email{julian.nickerl@uni-ulm.de}           %  \\
%             \emph{Present address:} of F. Author  %  if needed
}

\date{24 June 2019}
% The correct dates will be entered by the editor

\maketitle

\begin{abstract}
This work analyzes the minimum tollbooth problem in atomic network congestion games with unsplittable flows. The goal is to place tolls on edges, such that there exists a pure Nash equilibrium in the tolled game that is a social optimum in the untolled one. Additionally, we require the number of tolled edges to be the minimum. This problem has been extensively studied in non-atomic games, however, to the best of our knowledge, it has not been considered for atomic games before.
	
	By a reduction from the \wcnf{} problem, we show both the \np-hardness of the problem and the \w-hardness when parameterizing the problem with the number of tolled edges. On the positive side, we present a polynomial time algorithm for networks on series-parallel graphs that turns any given state of the untolled game into a pure Nash equilibrium of the tolled game with the minimum number of tolled edges.
% \PACS{PACS code1 \and PACS code2 \and more}
% \subclass{MSC code1 \and MSC code2 \and more}
\end{abstract}

\section{Introduction}

A class of games that are guaranteed to admit at least one pure Nash equilibrium are congestion games, as introduced by Rosenthal \cite{rosenthal1973class}. However, several well-known results (for example presented by Christodoulou and Koutsoupias \cite{christodoulou2005price} or Roughgarden and Tardos \cite{roughgarden2002bad}) indicate, that playing a Nash equilibrium rarely leads to a state that is socially beneficial. Especially in network congestion games, an approach to push the players towards a social optimum is the levy of tolls on edges of the network. The goal of these tolls is to turn a social optimum of the untolled game into a pure Nash equilibrium of the tolled game. The application of such tolls was shown to be highly beneficial (\cite{bilo2016dynamic,cole2006much}). If we allow the number of players to be infinite, each routing an infinitesimal amount of flow through the network, it is well known that tolls fulfilling the property exist and can be efficiently computed (\cite{beckmann1956studies,cole2003pricing,fleischer2004tolls,pigou1932economics}).

Games with a finite number of players are called atomic. The use of tollbooths in atomic games has a comparably short history. One of the main differences to the previous problem is that the strategy change of a single player has a substantial impact on all other players. Additionally, for the non-atomic games, it can be assumed that the resulting pure Nash equilibrium is unique, requiring only light restrictions, see, e.g., Roughgarden and Tardos \cite{roughgarden2002bad}. As shown by Orda, Rom, and Shimkin, \cite{orda1993competitive} this is not necessarily the case in atomic games.

Further differentiating the problem leads to the question: Does it suffice that there is at least one Nash equilibrium that corresponds to a social optimum, or must all Nash equilibria correspond to a social optimum?  Fotakis, Karakostas, and Kolliopoulos analyzed whether tolls fulfilling the desired properties exist. They conclude in \cite{fotakis2010existence} that in the latter case the existence cannot be guaranteed even for simple parallel-link networks. On the other hand, for the former case, they can confirm the existence of tolls in games where all players share the same source node and give a polynomial time algorithm calculating the tolls. Other work on this topic is, e.g., Fotakis and Spirakis \cite{fotakis2008cost} or Caragiannis, Kaklamanis, and Kanellopoulos \cite{caragiannis2006taxes}. Another line of research allows the players to split up their flow (\cite{cominetti2009impact,swamy2007effectiveness}). We focus on games with unsplittable flows; therefore we do not go into detail with this other branch.

Several papers have focused on special cases of the above problem, restricting, e.g., the height of the tolls (\cite{bonifaci2011efficiency}), the set of tollable edges (\cite{harks2015computing}), or knowledge about the number of players (\cite{colini2018demand}).
We consider here the minimum tollbooth problem, first introduced by Hearn and Ramana \cite{hearn1998solving}. In this problem, we additionally try to minimize the number of tolled edges. Work on this topic is mainly restricted to non-atomic games, with several heuristics as a result (e.g., \cite{bai2010heuristic,bai2009combinatorial,bai2008evolutionary,harwood2005genetic,stefanello2017minimization}). %\cite{buriol2010biased}
 The \np-hardness of the non-atomic case with multiple commodities was shown by Bai, Hearn, and Lawphongpanich in \cite{bai2010heuristic} and later for the single commodity case by Basu, Lianeas, and Nikolova in \cite{basu2015new}. On the positive side, Basu, Lianeas, and Nikolova give a polynomial time algorithm, if the network is a series-parallel graph \cite{basu2015new}.

\paragraph{Contribution:} In this work, we analyze the complexity of the minimum tollbooth problem for atomic network congestion games with unsplittable flows. As mentioned before, it is already known that such tolls may not exist considering the case in which every pure Nash equilibrium of the tolled game must correspond to a social optimum of the original one \cite{fotakis2010existence}. Therefore, we focus on the problem in which it is only required that there exist a pure Nash equilibrium that corresponds to a social optimum of the untolled game. By a reduction from the \wcnf{} problem, we show that the problem is \np-hard. Furthermore, we also prove that the parameterized problem considering the number of tollbooths as the parameter is \w-hard, giving evidence that the parameterized problem is not in \fpt. An additional spin-off result originates in the nature of the reduction: Finding a social optimum in atomic network congestion games with unsplittable flows is hard, supporting and extending existing results (see, e.g., Chakrabarty, Mehta, and Nagarajan \cite{chakrabarty2005fairness} or Meyers and Schulz\cite{meyers2008complexity}).

We also show that for a non-trivial graph class, it is possible to efficiently calculate a solution with the smallest possible number of toll booths. Based on the algorithm by Basu, Lianeas, and Nikolova from \cite{basu2015new}, we construct a polynomial time algorithm on games based on series-parallel graphs that turns any given state of the untolled game into a pure Nash equilibrium of the tolled game with the minimum number of tolled edges.

% Citations:
% Fotakis: Cost-balancing tolls for atomic network congestion games; Existence of cost balancing tolls, calculating "optimal" (with respect to the height of the costs) tolls is possible in linear time in series parallel graphs
% Fotakis: On the existence of Optimal Taxes for Network Congestion Games with Heterogeneous users; analyse heterogeneous users (different sensitivity to taxes). weakly optimal taxes always exist in single source games. strongly don't even exist in parallel link games
% Caragiannis: Taxes for linear atomic congestion games; How much can taxes improve the price of anarchy, also discuss existence of such toll vectors.
% Swamy: The effectiveness of stackelberg strategies and tolls for network congestion games; weakly optimal Tolls in games with splittable flows exist and can be computed in polynomial time

% Basu: New complexity results and Algorithms for the minimum tollbooth problem; some hardness results, P-Algorithm in series parallel networks.
% Hearn: Solving congestion toll pricing models; initial mention of the MINTB problem
% Corban:A genetic algorithm fot the minimum tollbooth problem; Bai: An evolutionary method for the minimum toll booth problem: The methodology; Buriol: A biased random-key genetic algorithm for road congestion minimization; heuristics for MINTB

% Give references specifically for MINTB

\section{Preliminaries}

This section introduces essential general definitions. In an attempt to shorten some statements, we will sometimes use the expression $[d]$ for the set $\{1,...,d\}$ for any natural number $d$ throughout the work.

\begin{definition}
	An \textit{atomic network congestion game} is a tuple
	\begin{center}
	$\Gamma = (G,N,(c_e)_{e\in E}, (s_i,t_i)_{i\in N})$
	\end{center}
	with $G = (V,E)$ being an undirected graph, $N = \{1,...,n\}$ a set of players, $c_{e}$ non-negative, monotonically increasing cost functions from the set of possible states of the game to the real numbers, and $(s_i,t_i)$ source-sink pairs with $s_{i},t_{i}\in V$. The strategies for a player $i$ are the paths from $s_{i}$ to $t_{i}$. The game is called \textit{symmetric}, if $s_{i} = s_j$ and $t_{i} = t_{j}$ for all players $i$ and $j$. It is called \textit{single-source} (\textit{single-sink}) if all players share the same source (sink). A state $S = (S_1,...,S_n)$ is a vector of strategies of players ($S_{i}$ is a strategy of player $i$). The congestion $n_{e}(S)$ on an edge $e$ is the number of players using that edge in their strategy in state $S$. The cost of a player in state $S$ is $\gamma_i(S) = \sum_{e \in S_{i}}c_{e}(n_{e}(S))$, where $S_{i}$ is the strategy of player $i$ in $S$. The \textit{social cost} of a state is the sum of the costs of all players. A \textit{social optimum} is a state with minimum social cost. A state $S$ is a pure Nash equilibrium, if for every player $i$ and every state $S'$, $\gamma_i(S) \leq \gamma_i((S'_{i},S_{-i}))$. Hereby, the state $(S'_{i},S_{-i})$ denotes player $i$ playing his strategy from $S'$ and all other players remaining on their strategy of state $S$.
\end{definition}

\begin{definition}
	The \textit{minimum tollbooth problem} (\mintb) on atomic network congestion games is the task of finding tolls for the edges of the network such that a social optimum of the untolled game is a pure Nash equilibrium in the tolled game while tolling only the minimum necessary number of edges. The tolls are depicted by a toll vector $\theta = (t_{e_1},...,t_{e_{\vert E\vert}})$, with $t_{i}\in \mathbb{R}^+_0$. An edge $e_{i}$ is called \textit{tolled} if $t_{e_{i}} > 0$, and \textit{untolled} otherwise.
	
	We say a state $S$ is implemented in an atomic network congestion game $\Gamma^\theta$ if $\theta$ is a toll vector such that $\Gamma^\theta = (G,N,(c_e^\theta)_{e\in E},(s_i,t_i)_{i\in{N}})$ with $c_e^\theta(S) = c_e(S) + t_{e}$ has $S$ as a pure Nash equilibrium. The state is optimally implemented if for this purpose the minimum necessary number of edges is tolled. 
\end{definition}

%Fotakis, Karakostas, and Kolliopoulos \cite{fotakis2010existence} guarantee the existence of toll vectors such that any state can be implemented in an atomic single-source network congestion game.

%\begin{theorem} \cite{fotakis2010existence}
%\label{theorem:existence}
%	Given a state $S$ and a single-source atomic network congestion game $\Gamma$, there exists a toll vector $\theta$, such that $S$ is a pure Nash equilibrium in $\Gamma^\theta$.
%\end{theorem}

%Their result is linked to a polynomial time algorithm that finds tolls that implement a state $S$ in the game. However, they require more than the minimal number of toll booths in general.
\section{Atomic \mintb{} is Hard}

This section gives the reduction of the \wcnf{} problem to the atomic \mintb{} problem. We consider formulas in conjunctive normal form (\cnf).

\begin{definition}
	A Boolean formula $F$ is in \cnf{} if $F = C_1\wedge ... \wedge C_n$. Hereby, $C_i$ is called clause and must have the form $C_i = (u_{i,1}\vee...\vee u_{i,k_{i}})$. The $u_{i,j}$ are called literals, with $u_{i,j}\in \{x_{1},...,x_{n}\}\cup\{\overline{x_{1}},...,\overline{x_{n}}\}$. The $x_{i}$ are called variables, with $x_{i}\in\{0,1\}$.
\end{definition}

\begin{definition} \wcnf{}:

	\begin{tabular}{ll}
	\textit{Input:} & A Boolean formula $F$ in \cnf{}\\
	\textit{Parameter:} & A positive integer $k$\\
	\textit{Question:} & Does $F$ have a satisfying assignment with weight $k$?
	\end{tabular}
	
	%\vspace*{2mm}
	The weight of an assignment is its Hamming weight, i.e., the number of variables with value 1.
\end{definition}

It is well known, that \wcnf{} is a \w-complete problem, see, e.g., chapter 12 of \cite{downey2012parameterized} or chapter 13 of \cite{cygan2015parameterized}. This gives evidence that the problem does not lie in the class \fpt.

\begin{definition}
	Let $\mathcal A$ be a parameterized problem with parameter $k$. We say that $\mathcal A$ is fixed-parameter tractable (\fpt), if there exists an algorithm solving $\mathcal A$ with running time in $\mathcal{O}(p(n)\cdot f(k))$. Here, $p(n)$ is a polynomial that depends on the problem size of $\mathcal A$ but not on the parameter $k$, and $f(k)$ is any function that only depends on the parameter $k$.	
\end{definition}

For more information on parameterized problems, we highly recommend \cite{downey2012parameterized} by Downey and Fellows or \cite{cygan2015parameterized} by Cygan et al.

We will reduce \wcnf{} to atomic \mintb{} in such a way, that the minimum number of necessary tollbooths is exactly the minimum weight of a satisfying assignment of the formula. We will construct an atomic network congestion game $\Gamma_{F}$ based on formula $F$, where $F$ is in \cnf{} with $n$ variables $x_1,...,x_n$ and $m$ clauses $C_{1},...,C_{m}$. We will call $\Lambda^0_{x_i}$ ($\Lambda^1_{x_i}$) the number of occurrences of variable $x_{i}$ in $F$ in negative (positive) form, $\Lambda_{x_i} = \Lambda^0_{x_i} + \Lambda^1_{x_i}$ and $\Lambda = \sum_{i=1}^n\Lambda_{x_i}$. The game will be played by a total of $m + \Lambda + n$ players, each fulfilling a specific role. The $m$ players, called \textit{clause players}, will each try to find a strategy that can be identified as satisfying a different clause. This will be done by each player choosing the occurrence of a literal in the clause in order to satisfy it. The $\Lambda$ players, or \textit{occurrence players}, will ensure that while satisfying the clauses, a variable cannot be assigned both value 0 and 1. Lastly, the $n$ players are the \textit{variable players}, that will state which edges have to be tolled to optimally implement a social optimum. Most players will have both different sources and different sinks.

For the reduction to be correct, the game $\Gamma_{F}$ has to fulfill the following properties:

\begin{enumerate}
\item[I)] Each \cp{} chooses a strategy that represents the satisfaction of a different clause.
\item[II)] If two \cp s choose the same variable as a representative, they both choose either its positive or its negative form, but not a mixture.
\item[III)] Optimally implementing a social optimum requires at least $k$ toll stations, where $k$ is the weight of the minimum satisfying assignment of $F$.
\item[IV)] From the tolled edges, the minimum satisfying assignment can be reconstructed in polynomial time.
\end{enumerate}

We start by constructing a graph for a game that satisfies property I). Only the \cp s play this game.

Let $G_1(V_1,E_1)$ be the undirected graph with

$\begin{array}{rcl}
V_1 &=& \{s\}\\
 & \cup & \{v_i,v_i^0,v_i^1 | i \in [n]\}\\
 
& \cup & \{l^0_{x_i,j,k},r^0_{x_i,j,k},z^0_{x_{i},j} | i \in [n],j \in [\Lambda^0_{x_i}], k \in[\Lambda^1_{x_i}] \}\\

&\cup&\{l^1_{x_i,j,k},r^1_{x_i,j,k},z^1_{x_{i},j}| i \in [n],j \in [\Lambda^1_{x_i}], k \in [\Lambda^0_{x_i}]\}\\

&\cup&\{c_i | i\in [m]\},\\

E_1 &=& \{\{s,v_i\},\{v_i,v_i^0\},\{v_i,v_i^1\} | i \in [n]\}\\

& \cup & \{\{v_i^0,l^0_{x_i,j,1}\},\{l^0_{x_i,j,k},r^0_{x_i,j,k}\},\{r^0_{x_i,j,k},l^0_{x_i,j,k+1}\} | i \in [n],j \in [\Lambda^0_{x_i}],\\
&& \hspace*{7cm}k \in[\Lambda^1_{x_i}]\}\\

& \cup & \{\{v_i^1,l^1_{x_i,j,1}\},\{l^1_{x_i,j,k},r^1_{x_i,j,k}\},\{r^1_{x_i,j,k},l^1_{x_i,j,k+1}\} | i \in [n],j \in [\Lambda^1_{x_i}],\\
&& \hspace*{7cm}k \in[\Lambda^0_{x_i}]\}\\

& \cup & \{\{r^0_{x_i,j,\Lambda^1_{x_i}},z_{x_{i},j}\},\{z_{x_{i},j},c_k\} | i \in [n],j \in [\Lambda^0_{x_i}],\\
&& \hspace*{4.25cm}k\mbox{ is the }j\mbox{th smallest index s.t. }\overline{x_{i}}\in C_{k}\}\\

& \cup & \{\{r^1_{x_i,j,\Lambda^0_{x_i}},z_{x_{i},j}\},\{z_{x_{i},j},c_k\} | i \in [n],j \in [\Lambda^1_{x_i}],\\
&& \hspace*{4.25cm}k\mbox{ is the }j\mbox{th smallest index s.t. }x_{i}\in C_{k}\}.
\end{array}$

The cost functions for the edges are a constant $7$ for each $\{v_i,v_i^1\}$ edge, a constant $2$ for each $\{v_{i},v_i^0\}$ edge and a constant $0$ for each $\{s,v_{i}\}$ edge. All other edges have cost $0$ for one, and $\infty$ (symbolizing a very large number) for any other number of players. As mentioned before, the game is played by the $m$ \cp s. Each \cp{} $i$ has $s$ as source and $c_i$ as sink.
Clearly, every node $c_{i}$ is reached by precisely one player, representing the satisfaction of property I). This is forced by every player having a different clause as their sink. Property I) could have been satisfied with a far simpler structure; however, this complex construction is necessary to fulfill the other properties in the upcoming steps. Also, the reason behind the choice of the cost functions will shortly become clear.

\begin{figure}[h]
%\vspace{-7mm}
\centering
\begin{tikzpicture}
	\node[circle,draw] (s) at (0,0){$s$};
	
	\node[circle,draw] (a) at (1,1.25){$A$};
	\node[circle,draw] (a0) at (2,0.5){$A^0$};
	\node[circle,draw] (a1) at (2,1.5){$A^1$};
	
	\node[circle,draw] (b) at (1,-1.25){$B$};
	\node[circle,draw] (b0) at (2,-1.5){$B^0$};
	\node[circle,draw] (b1) at (2,-0.5){$B^1$};

	\node[circle,draw] (la111) at (4,1.5){$l$};
	\node[circle,draw] (ra111) at (6,1.5){$r$};
	\node[circle,draw] (za111) at (8,1.5){$z$};
	
	\node[circle,draw] (la011) at (4,0.5){$l$};
	\node[circle,draw] (ra011) at (6,0.5){$r$};
	\node[circle,draw] (za011) at (8,0.5){$z$};
	
	\node[circle,draw] (lb111) at (4,-0.5){$l$};
	\node[circle,draw] (rb111) at (6,-0.5){$r$};
	\node[circle,draw] (zb111) at (8,-0.5){$z$};
	
	\node (C1) at (10,1.5){$\{A\}$};
	\node (C2) at (10,-0){$\{\overline A,B\}$};

	\path (s) -- node[left, xshift=-1mm]{$0$}(a);
	\draw (s) -- node[left]{$0$}(b);
	
	\path
	(a) -- node[above]{$7$}(a1) -- node[above]{$0|\infty$}(la111) -- node[above]{$0|\infty$}(ra111) -- node[above]{$0|\infty$}(za111) -- node[above]{$0|\infty$}(C1)
	(a) -- node[below]{$2$}(a0) -- node[above]{$0|\infty$}(la011) -- node[above]{$0|\infty$}(ra011) -- node[above]{$0|\infty$}(za011) -- node[above]{$0|\infty$}(C2);
	
	\path[draw]
	(b) -- node[above]{$7$}(b1) -- node[below]{$0|\infty$}(lb111) -- node[below]{$0|\infty$}(rb111) -- node[below]{$0|\infty$}(zb111) -- node[below]{$0|\infty$}(C2)
	(b) -- node[below]{$2$}(b0);

	\draw[red,->,ultra thick] (s) to[bend left=10] (a);
	\draw[red,->,ultra thick] (a) -- (a1);
	\draw[red,->,ultra thick] (a1) -- (la111);
	\draw[red,->,ultra thick] (la111) -- (ra111);
	\draw[red,->,ultra thick] (ra111) -- (za111);
	\draw[red,->,ultra thick] (za111) -- (C1);
	
	\draw[red,->,ultra thick] (s) to[bend right=10] (a);
	\draw[red,->,ultra thick] (a) -- (a0);
	\draw[red,->,ultra thick] (a0) -- (la011);
	\draw[red,->,ultra thick] (la011) -- (ra011);
	\draw[red,->,ultra thick] (ra011) -- (za011);
	\draw[red,->,ultra thick] (za011) -- (C2);
	
\end{tikzpicture}
%\vspace{-4mm}
\caption{The game on graph $G_{1}$ of formula $F = A\wedge(\overline A\vee B)$ played by two clause players. The red arrows indicate the social optimum of the game. Variable $A$ is assigned both $0$ and $1$ in the social optimum.}
\label{fig:propI}
%\vspace{-5mm}
\end{figure}
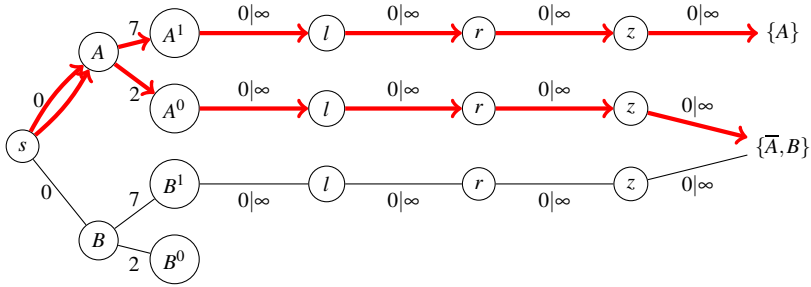

Figure \ref{fig:propI} displays this construction for the simple satisfiable formula $F = A\wedge(\overline A\vee B)$. For simplification, the names of the $l,r$ and $z$ nodes are shortened.

From now on, we call a path from a node $x_{i}^1$ to $z_{x_i}^1$ a 1-Path, and from node  $x_{i}^0$ to $z_{x_i}^0$ a 0-path. Similarly, we say a player assigns $1$ ($0$) to variable $x_{i}$ if he chooses the respective 1-Path (0-Path).

A path from a player from $s$ to an (unsatisfied) clause node $c_{j}$ chooses the positive or negative occurrence of a variable in $C_{j}$ in order to satisfy the clause. However, currently, it is possible that two clauses are reached over the same variable, while once in its negative and once in its positive form, i.e., property II) is not satisfied. This is the case in the example of Fig. \ref{fig:propI}. The red lines indicate the social optimum. The social optimum satisfies clause $\{\overline A,B\}$ by assigning $0$ to $A$ even though 1 was already assigned to $A$. We extend our game to fulfill property II). In addition to the \cp s, the game will now be played by the \op s as well.

Let $G_2(V_2,E_2)$ be the undirected graph with

$\begin{array}{rcl}
V_2 & = & V_1 \\
 & \cup & \{o^0_{i,j} | i \in [n],j \in [\Lambda_{x_i}^0]\}\\
 & \cup & \{o^1_{i,j} | i \in [n],j \in [\Lambda_{x_i}^1]\}\\
E_2 & = & E_1 \\
 & \cup & \{\{o^0_{x_i,j},z^0_{x_i,j}\} | i \in [n],j \in [\Lambda^0_{x_i}]\}\\
 & \cup & \{\{o^0_{x_i,j},l^1_{x_i,1,j}\} | i\in [n],j \in [\Lambda^0_{x_i}]\}\\
 
 & \cup & \{\{o^1_{x_i,j},z^1_{x_i,j}\} | i \in [n],j \in [\Lambda^1_{x_i}]\}\\
 & \cup & \{\{o^1_{x_i,j},l^0_{x_i,1,j}\} | i\in [n],j \in [\Lambda^1_{x_i}]\}\\
 
 & \cup & \{\{r^0_{x_i,j,k},l^0_{x_i,j+1,k}\} | i \in [n],j \in [\Lambda^0_{x_i}], k \in[\Lambda^1_{x_i}]\}\\
 & \cup & \{\{r^1_{x_i,j,k},l^1_{x_i,j+1,k}\} | i \in [n],j \in [\Lambda^1_{x_i}], k \in[\Lambda^0_{x_i}]\}\\
 
 & \cup & \{\{r^0_{x_i,\Lambda^0_{x_i},k},c_j\} | i \in [n], k \in[\Lambda^1_{x_i}],j\mbox{ is the }k\mbox{th smallest index s.t. } x_i \in C_j\}\\
 & \cup & \{\{r^1_{x_i,\Lambda^1_{x_i},k},c_j\} | i \in [n], k \in[\Lambda^0_{x_i}],j\mbox{ is the }k\mbox{th smallest index s.t. } \overline{x_i} \in C_j\}.
\end{array}$

The costs of edges from $E_{1}$ remain unchanged. The new edges $\{r_{x_i,j,k},c_l\}$, $\{o^1_{x_i,j},z^1_{x_i,j}\}$, and $\{o^0_{x_i,j},z^0_{x_i,j}\}$ have cost $6$ for one player and $\infty$ otherwise. All other edges have cost $0$ for one player and $\infty$ for any other number of players. The cost of $6$ ensures that a \cp{} cannot use the edge as a shortcut: Without using the new edge, his maximal cost is $7$, and the minimum cost with the edge is at least $8$. A total of $m + \Lambda$ players play the game on this graph. Each of the $\Lambda$ players represents the occurrence of a variable $x_i$ in clause $C_j$. If the occurrence is positive, the player has the source $o^1_{x_i,j}$, and $o^0_{x_i,j}$ if the occurrence is negative. In both cases, the sink is $c_j$.

\begin{figure}[h]
\centering
\begin{tikzpicture}
	\node[circle,draw] (s) at (1,0){$s$};
	
	\node[circle,draw] (A) at (2,0){$A$};
	\node[circle,draw] (A1) at (3,0.75){$A^1$};
	\node[circle,draw] (A0) at (3,-0.75){$A^0$};
	
	\node[draw] (o1a) at (8,1.5){$o^1_{A,1}$};
	\node[draw] (o1b) at (4,0.1){$o^1_{A,1}$};
	
	\node[draw] (o02a) at (6,1.5){$o^0_{A,2}$};
	\node[draw] (o02b) at (6,-1.4){$o^0_{A,2}$};
	
	\node[draw] (o01a) at (4,1.5){$o^0_{A,1}$};
	\node[draw] (o01b) at (6,0.1){$o^0_{A,1}$};
	
	\node[circle,draw] (1l1) at (4,0.75){$l$};
	\node[circle,draw] (1r1) at (5,0.75){$r$};
	\node[circle,draw] (1l2) at (6,0.75){$l$};
	\node[circle,draw] (1r2) at (7,0.75){$r$};
	\node[circle,draw] (1z) at (8,0.75){$z$};
	
	\node[circle,draw] (0l1) at (4,-0.75){$l$};
	\node[circle,draw] (0r1) at (5,-0.75){$r$};
	\node[circle,draw] (0z1) at (6,-0.75){$z$};
	
	\node[circle,draw] (0l2) at (4,-2.25){$l$};
	\node[circle,draw] (0r2) at (5,-2.25){$r$};
	\node[circle,draw] (0z2) at (6,-2.25){$z$};
	
	\node[draw] (c1) at (9,0.75){$c_1$};
	\node (c1b) at (10.2,0.75){$\{A\}$};
	\node[draw] (c1a) at (5,-3){$c_1$};
	
	\node[draw] (c2) at (9,-0.75){$c_2$};
	\node (c2b) at (10.2,-0.75){$\{\overline A,B\}$};
	\node[draw] (c2a) at (5,0){$c_2$};
	
	\node[draw] (c3) at (9,-2.25){$c_3$};
	\node (c3b) at (10.2,-2.25){$\{\overline A,\overline B,\overline C\}$};
	\node[draw] (c3a) at (7,0){$c_3$};

	% red path
	\draw[red,->,ultra thick] (s) -- (A);
	\draw[red,->,ultra thick] (A) -- (A1);
	\draw[red,->,ultra thick] (A1) -- (1l1);
	\draw[red,->,ultra thick] (1l1) -- (1r1);
	\draw[red,->,ultra thick] (1r1) -- (1l2);
	\draw[red,->,ultra thick] (1l2) -- (1r2);
	\draw[red,->,ultra thick] (1r2) -- (1z);
	\draw[red,->,ultra thick] (1z) -- (c1);

	% green path 1
	\draw[->,green,ultra thick] (o1b) -- (0l1);
	\draw[->,green,ultra thick] (0l1) -- (0r1);
	\draw[->,green,ultra thick] (0r1) -- (0l2);
	\draw[->,green,ultra thick] (0l2) -- (0r2);
	\draw[->,green,ultra thick] (0r2) -- (c1a);

	% green path 2
	\draw[->,green,ultra thick] (o01b) -- (0z1);
	\draw[->,green,ultra thick] (0z1) -- (c2);

	% green path 3
	\draw[->,green,ultra thick] (o02b) -- (0z2);
	\draw[->,green,ultra thick] (0z2) -- (c3);

	% black edges
	
	\draw (A) -- (A0);
	\draw (A0) -- (0l1);
	\draw (A0) -- (0l2);
	
	\draw (0r1) -- (0z1);
	\draw (0r2) -- (0z2);
	\draw (1r1) -- (c2a);
	\draw (1r2) -- (c3a);
	
	\draw (o1a) -- (1z);
	\draw (o01a) -- (1l1);
	\draw (o02a) -- (1l2);

\end{tikzpicture}
%\vspace{-4mm}
\caption{The segment of the graph $G_{2}$ of formula $F = A\wedge(\overline A\vee B)\wedge(\overline A\vee\overline B\vee\overline C)$ concerning variable $A$. Depicted are the strategies of one clause player (red arrows) and three occurrence players (green arrows) in the social optimum.}
\label{fig:propII}
%\vspace{-5mm}
\end{figure}
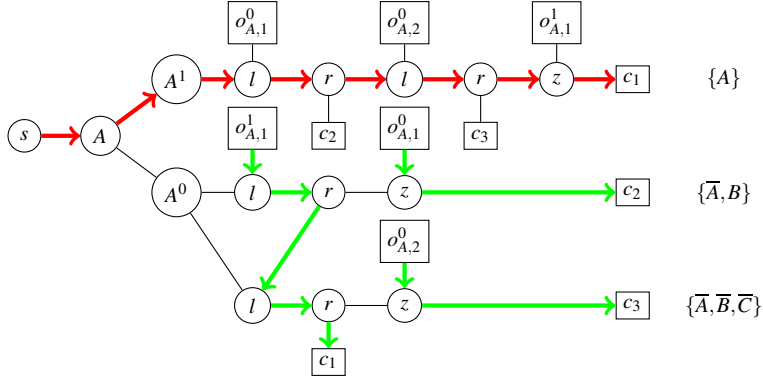

The graph in Fig. \ref{fig:propII} is the result of creating the construction for the formula $F = A\wedge(\overline A\vee B)\wedge(\overline A\vee\overline B\vee\overline C)$, while focusing only on the section concerning variable $A$. In an attempt to make the graph clearer, some nodes appear multiple times in the graph; however they all represent one single node. Nodes with this property are displayed as rectangles. If they have the same name, they represent the same single node.

The red path indicates a \cp{} satisfying $C_1$ by assigning 1 to $A$. The green edges encompass three \op s, each trying to reach another clause. The intention of this construction is to fix an assignment to a variable. If not hindered, an \op{} will try to reach a clause by going straight to the $z$ node. However, if there is a player that wants to satisfy the corresponding clause (e.g., by assigning the value 1 to it), the edge $z-C$ cannot be used by the \op{} as well. Instead, he takes the detour via the zigzag path, blocking all 0-paths of this variable, therefore satisfying property II).

A last problem with the construction is that the desired state reached when all players follow their intended strategies is a pure Nash equilibrium. Since the state is a Nash equilibrium already, no tolls are necessary to implement the social optimum, hence violating property III). A last small change solves this problem. From now on, the \vp s join the game.

Let $G_3(V_2,E_3)$ be the undirected graph with

$\begin{array}{rcl}
E_3 & = & E_2\ \cup\  \{\{v_{i}^0,v_{i}^1\}| i \in [n]\}. 
\end{array}$

The cost function of all edges from $E_{2}$ remain the same. The cost function for a new edge $e = \{v_{i}^0,v_{i}^1\}$ is $c_{e}(x) = 2x$, where $x$ is the number of players that use edge $e$. Now all $m+\Lambda+n$ players take part in the game. A \vp{} $i$ has $v_{i}^0$ as source and $v_{i}^1$ as sink.

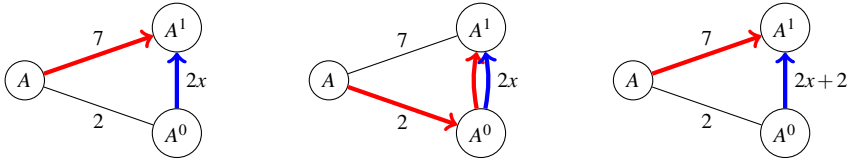
\begin{figure}
%\vspace{-7mm}
\centering
\begin{tikzpicture}
	\node[circle,draw] (v) at (0,0){$A$};
	\node[circle,draw] (v1) at (2,0.7){$A^1$};
	\node[circle,draw] (v0) at (2,-0.7){$A^0$};
	
	\draw[red,->,ultra thick] (v) --node[above,black]{$7$} (v1) ;
	\draw (v) --node[below,black]{$2$} (v0);
	\draw[blue,->,ultra thick] (v0) --node[right,black]{$2x$} (v1);
	
	\node[circle,draw] (vv) at (4,0){$A$};
	\node[circle,draw] (vv1) at (6,0.7){$A^1$};
	\node[circle,draw] (vv0) at (6,-0.7){$A^0$};
	
	\draw (vv) --node[above,black]{$7$} (vv1);
	\draw[red,->,ultra thick] (vv) to node[below,black]{$2$} (vv0);
	
	\draw[red,->,ultra thick] (vv0) to[bend left=10] (vv1);
	\draw[blue,->,ultra thick] (vv0) to[bend right=10]node[right,black]{$2x$} (vv1);
	
	\node[circle,draw] (vvv) at (8,0){$A$};
	\node[circle,draw] (vvv1) at (10,0.7){$A^1$};
	\node[circle,draw] (vvv0) at (10,-0.7){$A^0$};
	
	\draw[red,->,ultra thick] (vvv) --node[above,black]{$7$} (vvv1) ;
	\draw (vvv) --node[below,black]{$2$} (vvv0);
	\draw[blue,->,ultra thick] (vvv0) --node[right,black]{$2x+2$} (vvv1);
	
\end{tikzpicture}
%\vspace{-4mm}
\caption{A demonstration of the effect of the variable player (blue arrow). The clause player (red arrow) leaves the social optimum (left graph) for the pure Nash equilibrium (center graph). Tolling an edge prevents this defection (right graph).}
\label{fig:propIII}
%\vspace{-5mm}
\end{figure}

Figure \ref{fig:propIII} illustrates the new component.

The red edges correspond to a \cp{}, the blue edge to a \vp{}. The state on the left is the social optimum. However, it is not a pure Nash equilibrium, because the \cp{} can find a cheaper path, reducing his cost by 1 but also increasing the social cost by 1, displayed in the center state. The state on the right is a pure Nash equilibrium that coincides with the social optimum of the original graph. It is the result of tolling edge $\{A^0,A^1\}$ with 2.

To optimally implement the social optimum, placing this toll is necessary if and only if a \cp{} satisfies a clause by assigning 1 to a variable, and at most once per variable. Since assigning 1 to a variable is more expensive than assigning 0, the least number of variables will be assigned value 1, forcing the placement of the toll station. Tolling such an edge corresponds to assigning $1$ to the respective variable in the Boolean formula. Therefore, the minimal satisfying assignment for the original formula can be constructed from the placed toll stations efficiently. Additionally, if the social optimum can be optimally implemented with at most $k$ tollbooths, then there exists a satisfying assignment for the original formula with weight at most $k$. Graph $G_{3}$ therefore finally satisfies all four properties.
From these reductions follows Theorem \ref{th:hard}.

\begin{theorem}
\label{th:hard}
Solving atomic \mintb{} is \np-hard and \w-hard with the number of required toll booths as the parameter.
\end{theorem}

	To know where to place the tollbooths and therefore the minimal satisfying assignment of the original formula, it suffices to know the social optimum. A direct consequence is that finding a social optimum must be hard.
	
	\begin{corollary}
		Finding a social optimum of a game $\Gamma$ is \np-hard. Additionally, finding a social optimum that can be implemented in $\Gamma$ with at most $k$ tollbooths is \w-hard, with $k$ being the parameter.
	\end{corollary}

\section{Optimally Implementing a State in Polynomial Time}

This section presents an algorithm that optimally implements a state $S$ in an atomic network congestion game based on a series-parallel graph in polynomial time. We base the procedure on a similar approach from Basu, Lianeas, and Nikolova, who show the same result in \cite{basu2015new} for non-atomic games. Starting at the simple base case of parallel-link networks, we inductively decide on the number of tolled edges for larger components, based on the optimality of the smaller ones. We do so by exploiting the recursive structure of series-parallel graphs.

\begin{definition}
	A graph $G$ with source $s$ and sink $t$ is called \textit{series-parallel}, if
	\begin{enumerate}
		\item[i)] it consists of only a single edge.
		\item[ii)] it is the result of combining two series-parallel graphs in series.
		\item[iii)] it is the result of combining two series-parallel graphs in parallel.
	\end{enumerate}
	
	A combination of graphs $G_1$ and $G_2$ in series means declaring the source of $G_1$ as the new global source, and the sink of $G_2$ as the new global sink. Additionally, the sink of $G_1$ and the source of $G_2$ are identified as one node in the new graph.
	
	Combining $G_1$ and $G_2$ in parallel means identifying both sources as one node and setting it as the new global source, and respectively identifying both sinks as the new global sink.
	
	A \textit{series-parallel parse tree} of a series-parallel graph is the representation of the graph as a tree, where every leaf stands for a single edge, and every inner node represents either a combination of its two children in series or parallel. For simplicity, in this paper, the leaves will represent parallel-link networks, i.e., networks consisting of two nodes connected by possibly several edges.
\end{definition}

Given an atomic network congestion game on a series-parallel network, let $T$ be the parse tree representation of that network, and $S$ a given state that is to be implemented in that game. For each node of $T$, we will create a list of tuples $(\eta,\lambda)$ with the following meaning. While ensuring that all players follow their strategy in $S$, by tolling at most $\eta$ edges, $\lambda$ is the highest cost we can force on a new player entering the network represented by the regarded node. The values for $\eta$ range from the minimum necessary to guarantee that all player follow their strategy from $S$ to the minimum necessary to toll all paths through the network. Additionally, we will remember a single value $\lambda_{0}$, indicating the lowest cost a new player can have, while ensuring that $S$ is implemented.

Algorithm \ref{algo:PL} creates the list for a leaf of the tree, which is a parallel-link network. For simplicity, we denote by $l_{e} = c_{e}(n_{e}(S))$, and by $l_{e}^{+} = c_{e}(n_{e}(S)+1)$.
%\vspace{-7mm}
\begin{algorithm}[h]
\label{algo:PL}
\KwData{Graph $G = (V,E)$ represented by leaf $v$ of the parse tree, state $S$}
Set $list_v = []$\;
Set $l_{\max} = \max\{l_e |\ e\in E\}$\;
Set $\eta_{\min} = \vert\{e\ |\ e\in E\mbox{ and }l_e^+<l_{\max}\}\vert$\;
Set $\eta_{\max} = \vert E\vert$\;
Set $\lambda_{0} = \max\{\max\{l_e\},\min\{l_{e'}^+\}\ \vert\ e,e'\in E\}$

\For{$\eta = \eta_{\min}\mbox{ to }\eta_{\max}$}{
	Let $e_1,...,e_k$ be all edges, and $l_{1}^+\leq...\leq l_{k}^+$\\

	$\lambda = \infty$\;
	\If{$\eta\neq \eta_{\max}$}{
		$\lambda = l_{\eta+1}^+$\;
	}
	Append $(\eta,\lambda)$ to $list_v$\;
} 
\caption{ParallelLink}
\end{algorithm}
%\vspace{-7mm}
Lines 2 and 3 of Algorithm \ref{algo:PL} ensure, that at least the minimum number of edges necessary to implement the given state $S$ in the parallel-link network is tolled. Line 10 calculates the highest cost enforceable on a new player entering the network. Since the $\eta$ cheapest edges are tolled, the highest cost we could force on a newly entering player is the cost of the $\eta +1$ cheapest edge. Only when all edges are tolled, we can guarantee an arbitrarily high cost, marked by the $\infty$ symbol. Thus, the tuple $(\eta_{\max},\infty)$ forms the last element of the list.

Since the list is ordered with respect to the values of $\eta$, we will address elements from the list by index, so, e.g., $\eta_1$ refers to $\eta_{\min}$, and $\lambda_{i}$ is the cost enforceable by $\eta_{i}$ edges. Whenever we refer to the last element of a list, we use $\max$ as the index.

Following an inductive argument, we now show how to correctly form these lists for the inner nodes of the parse tree, assuming that the lists for the children are already correctly formed. To differentiate between the two children of the inner node $r$, we address one of them by $v$, and the other one by $w$. Some values will be labeled with $r,v$ or $w$ accordingly. A tuple for the list of $r$ will originate from two tuples, one for each child node. We add two pointers, $idx_{v}$ and $idx_w$, to each new tuple, referencing their origin.

\paragraph{Series composition:} The minimum number of edges that need to be tolled in the parent node $r$ is simply the sum of the minimum necessary number of edges from the children, so $\eta_{1}^r = \eta_{1}^v + \eta_{1}^w$. Similarly, the maximum $\eta_{\max^r}^r$ can be calculated by $\eta_{\max^r}^r = \min\{\eta_{1}^v + \eta_{\max^w}^w, \eta_{\max^v}^v + \eta_{1}^w\}$. To complete the list, we have to add a tuple for every $\eta_{i}^r$ with $1 < i < \max^r$. The corresponding costs are $\lambda_i^r = \max\{\lambda_a^v + \lambda_b^w\ \vert\ \eta_i^r = \eta_a^v + \eta_b^w\}$, for all $i\in[\max^r]$. Accordingly, we set $\lambda_{0}^r = \lambda_{0}^v + \lambda_{0}^w$.

%\vspace{-1mm}
\paragraph{Parallel composition:} In the parallel composition a player from one component may be able to reduce his cost by joining the other component. Let $c_{\max}$ be the highest cost of a player in both $v$ and $w$, if all players play according to the state $S$. Without loss of generality, let it be a player from $v$ that has this cost $c_{\max}$. To determine the minimum necessary number of tolled edges, it suffices to check how many edges in $w$ have to be tolled to force a joining player to have cost at least $c_{\max}$. That means, $\eta_1^r = \eta_1^v + \min\{ \eta^w_i\ \vert\ \lambda^w_i \geq c_{\max}, i > 0\}$, and $\eta_{\max^r}^r = \eta_{\max^v}^v + \eta_{\max^w}^w$. Again, we add a tuple for every $\eta_{i}^r$ with $1 < i < \max^r$. The corresponding costs are $\lambda_{i}^r = \max\{\min\{\lambda_a^v,\lambda_b^w\}\ \vert\ \eta_i^r = \eta_a^v + \eta_b^w\}$, for all $i \in [\max^r]$. We set $\lambda_{0}^r = \max\{c_{\max},\min\{\lambda_{0}^v,\lambda_{0}^w\}\}$.\\[2mm]
Once the lists for all nodes are created, we can choose the tuple in the root node with the minimum number of edges. We can retrace the creation of the tuple to the leaf nodes of the tree, where we can decide which edges are to be tolled with which value. Algorithm \ref{algo:place} is a recursive algorithm that tolls the edges accordingly. For simplicity, we assume the lists of each node to be globally accessible. The algorithm is initially called with $r$ being the root node of the parse tree and $c_{in} = \lambda_{0}^r$, since optimally implementing $S$ requires tolling $\eta_{1}^r$ edges.

\begin{algorithm}[h]
\label{algo:place}
\KwData{Parse Tree $T$, current node $r$, cost $c_{in}$}
\If{$r$ is a leaf node}{
	Let $(V,E)$ be the parallel-link network represented by $r$\;
	\For{each edge $e\in E$ with $l_e^+ < c_{in}$}{
		set toll $t_{e} = c_{in}-l_{e}^+$\;
	}
	\KwRet
}

Let $v = $ left child of $r$\;
Let $w = $ right child of $r$\;
Set $i = \arg\min_{i}\{\lambda_{i}^r\geq c_{in}\ \vert\ i > 0\}$\;
Let $j$ be the $idx_{v}$ of tuple $(\eta_i^r,\lambda_i^r)$\;
Let $k$ be the $idx_{w}$ of tuple $(\eta_i^r,\lambda_i^r)$\;

\eIf{$r$ is a series composition}{
	Set $c_{out}^v$ and $c_{out}^w$, s.t.:\\
	$\ -\ \ c_{out}^v + c_{out}^w = c_{in},$\\
	$\ -\ \ c_{out}^v > \lambda_{j-1}^v$ if $j > 1$, else $c_{out}^v \geq\lambda_{0}^v,$\\
	$\ -\ \ c_{out}^w > \lambda_{k-1}^w$ if $k > 1$, else $c_{out}^w \geq\lambda_{0}^w$\;
}
{
	Set $c_{out}^v = \max\{c_{in}, \lambda_{0}^v\}$\;
	Set $c_{out}^w = \max\{c_{in}, \lambda_{0}^w\}$\;
}

Call PlaceTolls$(T,v,c_{out}^v)$\;
Call PlaceTolls$(T,w,c_{out}^w)$\;

\caption{PlaceTolls}
\end{algorithm}

Lines $1$ to $5$ toll the edges in the parallel-link networks at the leaves of the parse tree. Lines $11$ to $15$ consider the case when $r$ is a series composition. Here the input cost is feasibly divided into two components. By this division, it is both possible and necessary to implement $S$ with the number of edges indicated by the $idx$ pointer. Lines $16$ to $18$ do the same for the parallel composition.  This way, it is always possible to toll the rest of the network feasibly while only using $\eta_{1}^r$. It ensures, that the state gets properly implemented in all components of the series-parallel graph, and a new player joining the network has cost at least the initial $c_{in}$.

%\vspace{-1mm}
\paragraph{Runtime:} Let $m$ be the number of edges in the original network. The parse tree can be created in $\mathcal{O}(m)$ time, and also has size $\mathcal{O}(m)$. Creating the list at a leaf node is possible in $\mathcal{O}(m)$ time. For an inner node, we have to check at most $\mathcal{O}(m^2)$ combinations of elements from the children's lists. Therefore, creating the list for every node is possible in $\mathcal{O}(m^3)$ time. The computation time on each node in Algorithm \ref{algo:place} is bounded by $\mathcal{O}(m)$. The algorithm visits every node of the parse tree exactly once, leading to a runtime of $\mathcal{O}(m^2)$. In total, the runtime of the algorithm is $\mathcal{O}(m^3)$. 

\begin{theorem}
	Optimally implementing a state $S$ in an atomic network congestion game based on a series-parallel graph with $m$ edges is possible in runtime $\mathcal{O}(m^3)$.
\end{theorem}

%\newpage
\section{Conclusion}

This work analyses the complexity of the minimum tollbooth problem in atomic network congestion games with unsplittable flows. 

By a reduction from the \wcnf{} problem, we show both the \np-hardness of the problem and, more importantly, the \w-hardness of a parameterized version. The parameter in the latter case is the number of necessary tollbooths to turn a social optimum of the untolled game into a pure Nash equilibrium of the tolled one.

A restriction to networks that are series-parallel graphs yields a polynomial time algorithm. The algorithm gives tolls, such that a given state is a pure Nash equilibrium of the tolled game while using the minimum number of toll booths. This algorithm is based on a similar method presented by Basu, Lianeas, and Nikolova in \cite{basu2015new}, who consider non-atomic games, exploiting the recursive structure of series-parallel graphs. 

% BibTeX users please use one of
%\bibliographystyle{spbasic}      % basic style, author-year citations
\bibliographystyle{spmpsci}      % mathematics and physical sciences
%\bibliographystyle{spphys}       % APS-like style for physics
%\bibliography{atomicMintb}   % name your BibTeX data base

\begin{thebibliography}{10}
\providecommand{\url}[1]{{#1}}
\providecommand{\urlprefix}{URL }
\expandafter\ifx\csname urlstyle\endcsname\relax
  \providecommand{\doi}[1]{DOI~\discretionary{}{}{}#1}\else
  \providecommand{\doi}{DOI~\discretionary{}{}{}\begingroup
  \urlstyle{rm}\Url}\fi

\bibitem{bai2010heuristic}
Bai, L., Hearn, D.W., Lawphongpanich, S.: A heuristic method for the minimum
  toll booth problem.
\newblock Journal of Global Optimization \textbf{48}(4), 533--548 (2010)

\bibitem{bai2009combinatorial}
Bai, L., Rubin, P.A.: Combinatorial benders cuts for the minimum tollbooth
  problem.
\newblock Operations research \textbf{57}(6), 1510--1522 (2009)

\bibitem{bai2008evolutionary}
Bai, L., Stamps, M.T., Harwood, R.C., Kollmann, C.J.: An evolutionary method
  for the minimum toll booth problem: The methodology.
\newblock Journal of Management Information and Decision Sciences  (2008)

\bibitem{basu2015new}
Basu, S., Lianeas, T., Nikolova, E.: New complexity results and algorithms for
  the minimum tollbooth problem.
\newblock In: International Conference on Web and Internet Economics, pp.
  89--103. Springer (2015)

\bibitem{beckmann1956studies}
Beckmann, M., McGuire, C.B., Winsten, C.B.: Studies in the economics of
  transportation.
\newblock Yale University Press  (1956)

\bibitem{bilo2016dynamic}
Bil{\`o}, V., Vinci, C.: Dynamic taxes for polynomial congestion games.
\newblock In: Proceedings of the 2016 ACM Conference on Economics and
  Computation, pp. 839--856. ACM (2016)

\bibitem{bonifaci2011efficiency}
Bonifaci, V., Salek, M., Sch{\"a}fer, G.: Efficiency of restricted tolls in
  non-atomic network routing games.
\newblock In: International Symposium on Algorithmic Game Theory, pp. 302--313.
  Springer (2011)

\bibitem{caragiannis2006taxes}
Caragiannis, I., Kaklamanis, C., Kanellopoulos, P.: Taxes for linear atomic
  congestion games.
\newblock In: ESA, pp. 184--195. Springer (2006)

\bibitem{chakrabarty2005fairness}
Chakrabarty, D., Mehta, A., Nagarajan, V.: Fairness and optimality in
  congestion games.
\newblock In: Proceedings of the 6th ACM Conference on Electronic Commerce, pp.
  52--57. ACM (2005)

\bibitem{christodoulou2005price}
Christodoulou, G., Koutsoupias, E.: The price of anarchy of finite congestion
  games.
\newblock In: Proceedings of the thirty-seventh annual ACM symposium on Theory
  of computing, pp. 67--73. ACM (2005)

\bibitem{cole2003pricing}
Cole, R., Dodis, Y., Roughgarden, T.: Pricing network edges for heterogeneous
  selfish users.
\newblock In: Proceedings of the thirty-fifth annual ACM symposium on Theory of
  computing, pp. 521--530. ACM (2003)

\bibitem{cole2006much}
Cole, R., Dodis, Y., Roughgarden, T.: How much can taxes help selfish routing?
\newblock Journal of Computer and System Sciences \textbf{72}(3), 444--467
  (2006)

\bibitem{colini2018demand}
Colini-Baldeschi, R., Klimm, M., Scarsini, M.: Demand-independent optimal tolls
   (2018)

\bibitem{cominetti2009impact}
Cominetti, R., Correa, J.R., Stier-Moses, N.E.: The impact of oligopolistic
  competition in networks.
\newblock Operations Research \textbf{57}(6), 1421--1437 (2009)

\bibitem{cygan2015parameterized}
Cygan, M., Fomin, F.V., Kowalik, {\L}., Lokshtanov, D., Marx, D., Pilipczuk,
  M., Pilipczuk, M., Saurabh, S.: Parameterized algorithms, vol.~3.
\newblock Springer (2015)

\bibitem{downey2012parameterized}
Downey, R.G., Fellows, M.R.: Parameterized complexity.
\newblock Springer (1999)

\bibitem{fleischer2004tolls}
Fleischer, L., Jain, K., Mahdian, M.: Tolls for heterogeneous selfish users in
  multicommodity networks and generalized congestion games.
\newblock In: Proceedings of the fourty-fifth annual IEEE symposium on
  Foundations of Computer Science, pp. 277--285. IEEE (2004)

\bibitem{fotakis2010existence}
Fotakis, D., Karakostas, G., Kolliopoulos, S.G.: On the existence of optimal
  taxes for network congestion games with heterogeneous users.
\newblock In: International Symposium on Algorithmic Game Theory, pp. 162--173.
  Springer (2010)

\bibitem{fotakis2008cost}
Fotakis, D., Spirakis, P.G.: Cost-balancing tolls for atomic network congestion
  games.
\newblock Internet Mathematics \textbf{5}(4), 343--363 (2008)

\bibitem{harks2015computing}
Harks, T., Kleinert, I., Klimm, M., M{\"o}hring, R.H.: Computing network tolls
  with support constraints.
\newblock Networks \textbf{65}(3), 262--285 (2015)

\bibitem{harwood2005genetic}
Harwood, R.C., Kollmann, C.J., Stamps, M.T.: A genetic algorithm for the
  minimum tollbooth problem  (2005)

\bibitem{hearn1998solving}
Hearn, D.W., Ramana, M.V.: Solving congestion toll pricing models.
\newblock In: Equilibrium and advanced transportation modelling, pp. 109--124.
  Springer (1998)

\bibitem{meyers2008complexity}
Meyers, C.A., Schulz, A.S.: The complexity of congestion games.
\newblock Massachusetts Institute of Technology, Cambridge pp. 1--16 (2008)

\bibitem{orda1993competitive}
Orda, A., Rom, R., Shimkin, N.: Competitive routing in multiuser communication
  networks.
\newblock IEEE/ACM Transactions on Networking (ToN) \textbf{1}(5), 510--521
  (1993)

\bibitem{pigou1932economics}
Pigou, A.C.: The economics of welfare.
\newblock McMillan\&Co., London  (1920)

\bibitem{rosenthal1973class}
Rosenthal, R.W.: A class of games possessing pure-strategy nash equilibria.
\newblock International Journal of Game Theory \textbf{2}(1), 65--67 (1973)

\bibitem{roughgarden2002bad}
Roughgarden, T., Tardos, {\'E}.: How bad is selfish routing?
\newblock Journal of the ACM (JACM) \textbf{49}(2), 236--259 (2002)

\bibitem{stefanello2017minimization}
Stefanello, F., Buriol, L.S., Hirsch, M.J., Pardalos, P.M., Querido, T.,
  Resende, M.G., Ritt, M.: On the minimization of traffic congestion in road
  networks with tolls.
\newblock Annals of Operations Research \textbf{249}(1-2), 119--139 (2015)

\bibitem{swamy2007effectiveness}
Swamy, C.: The effectiveness of stackelberg strategies and tolls for network
  congestion games.
\newblock In: Proceedings of the eighteenth annual ACM-SIAM symposium on
  Discrete algorithms, pp. 1133--1142. Society for Industrial and Applied
  Mathematics (2007)

\end{thebibliography}

% Non-BibTeX users please use
%\begin{thebibliography}{}
%
% and use \bibitem to create references. Consult the Instructions
% for authors for reference list style.
%
%\bibitem{RefJ}
% Format for Journal Reference
%Author, Article title, Journal, Volume, page numbers (year)
% Format for books
%\bibitem{RefB}
%Author, Book title, page numbers. Publisher, place (year)
% etc
%\end{thebibliography}

\end{document}